# Expeditious stochastic approach for MP2 energies in large electronic systems

Daniel Neuhauser[†], Eran Rabani[‡] and Roi Baer[♣]

[†]Department of Chemistry and Biochemistry, University of California, Los Angeles CA-90095, USA.
[‡]School of Chemistry, The Sackler Faculty of Exact Sciences, Tel Aviv University, Tel Aviv 69978, Israel.
[♣]Fritz Haber Center for Molecular Dynamics, Institute of Chemistry, Hebrew University, Jerusalem 91904 Israel.

**ABSTRACT**: A fast stochastic method for calculating the 2$^{nd}$ order Møller-Plesset (MP2) correction to the correlation energy of large systems of electrons is presented. The approach is based on reducing the exact summation over occupied and unoccupied states to a time-dependent trace formula amenable to stochastic sampling. We demonstrate the abilities of the method to treat systems of thousands electrons using hydrogen passivated silicon spherical nanocrystals represented on a real space grids, much beyond capabilities of present day MP2 implementations.

Post Hartree-Fock (HF), 5th rung density functional theory (DFT) and many-body perturbation theory (MBPT) calculations are of importance for estimation of observables dependent on correlation energy in large systems. Examples include the calculation of cohesion energies and molecular geometries, phonon/vibrational properties in large molecules,[1-11] electron dynamics,[12-15] and quasiparticle energies and gaps in nanocrystals.[16-23] Application of such methods in a straightforward way to large systems of experimental relevance is often hampered by the steep scaling of the computational effort (CPU time and memory) with system size. This is mainly due to the huge number of electron-electron Coulomb integrals needed for the calculation, namely between all pairs of occupied and un-occupied orbitals. The problem is especially critical for grid-based calculations, where the number of unoccupied orbitals easily reaches millions. Thus, accurate first principles post HF/DFT methods cannot treat but the smallest systems and there is an urgent need for new ideas to overcome the formidable barriers.

Lowering the scaling of such calculations is the topic of recent works, such as localized density fitting methods[24,25] and virtual orbital selection techniques.[26-28] Such approaches lead to the development of linear scaling local 2$^{nd}$ order Møller-Plesset (LMP2) techniques[11,24] which are extremely efficient for low dimensional systems and have been used successfully for small molecules containing $O(100)$ atoms. However, for higher dimensional systems, these methods are too expensive due to the computational cost of a prefactor.

In this letter, we develop an alternative approach that allows treating post HF/DFT calculations for large systems and is not particularly sensitive to dimensionality. The basic idea is to use stochastic methods, giving up accuracy (introducing statistical errors) in exchange for efficiency in a controlled and automated way. Ultimately, we aim at systems such as nanocrystals, containing $O(10^4)$ atoms and $O(10^5)$ valence electrons where one does not need high accuracy in absolute total energy but requires a small average error per electron. As an example of a case of interest, we refer the reader to a recent stochastic correlation method that has been developed to study multiexciton generation (MEG) rates in nanocrystals.[29] The calculation can address questions such as the dependence of the MEG rate on the size and shape of the nanocrystal or the comparison of the MEG rate with phonon emission decay processes with relative accuracies of 1% at reasonable computational effort

Here, we focus on the simplest of correlated electron calculations, namely the MP2 correlation energy, yet the concept we introduce is general and transferable to more challenging correlation calculations. We rely on a novel combination of stochastic and operator techniques allowing MP2 calculations to be performed on grid-based representation with linear scaling effort both in time and storage. The methods we develop here share some ideas from recent work[29-31] but the present combination and its use for MP2 is new.

The MP2 energy for a closed shell system, in terms of spatial orbitals, is:

$$E_{MP2} = \sum_{ijab} \frac{(2\langle ab|ij\rangle - \langle ab|ji\rangle)\langle ij|ab\rangle}{\epsilon_i + \epsilon_j - \epsilon_a - \epsilon_b}, \qquad (1)$$

Here, $\langle ab|ij\rangle = \iint a(x_1)^* b(x_2)^* x_{12}^{-1} i(x_1) j(x_2) d^3x_1 d^3x_2$, $\epsilon_i$, $\epsilon_j$ ($\epsilon_a$, $\epsilon_b$) and $i(x)$, $j(x)$ ($a(x)$, $b(x)$) are the occupied (virtual) eigenvalues and eigenfunctions of the Fock operator $H_0$. When calculated as-is without further processing, we label this formula as "full-summation". For nanocrystals of experimentally relevant sizes the density of states in the valence and conduction bands reaches several thousands per eV,[29] thus the summation in Eq. (1) (2) involves more than $10^{12}$ terms but the non-local nature of the orbitals increases the computations to $10^{15} - 10^{16}$ operations.

To replace the full summation by random averages we need to circumvent the state-specific energy denominator. This is done through time integration, using the identity $\text{Im} \int_0^\infty e^{i\epsilon t} \text{erfc}\left(\frac{t}{\tau}\right) dt = \frac{1}{\epsilon}\left(1 - e^{-\left(\frac{\epsilon\tau}{2}\right)^2}\right)$, where $\text{erfc}(x)$ is the complementary error function. This expression converges to $\frac{1}{\epsilon}$ once $\tau \gg \frac{1}{2E_g}$ where $E_g$ is the HOMO-LUMO gap (note: one can also use the Laplace transform[32] instead of the above identity. The advantage of using the erfc is that it is not sensitive to the sign of $\epsilon$ and so can be used for cases where denominators are not necessarily positive). Defining $\phi_i(t) \equiv e^{-i\epsilon_i t} j(x) = (e^{-iH_0 t} i)(x)$ and (we set $\hbar = 1$), the MP2 correlation energy takes the form:

$$E_{MP2} = \text{Im} \int_0^\infty C(t)\, \text{erfc}\left(\frac{t}{\tau}\right) dt, \qquad (2)$$

where:

$$C(t) \equiv \sum_{ijab} (2\langle ab|ij\rangle - \langle ab|ji\rangle) \times \langle i(t)j(t)|a(t)b(t)\rangle. \qquad (3)$$

We now replace the summation by stochastic averaging procedure. We introduce four random wavefunctions:

$$\eta(x) \equiv \sum_i \eta_i i(x), \quad \chi(x) \equiv \sum_j \chi_j j(x),$$
$$\xi(x) \equiv \sum_a \xi_a a(x), \quad \zeta(x) \equiv \sum_b \zeta_b b(x) \qquad (4)$$

where $\eta_i$, etc., are independent complex random numbers, normalized so that $[\eta_i^* \eta_{i'}] = \delta_{ii'}$ (and similarly for the $\chi$'s, $\zeta$'s and $\xi$'s) and $[\eta_i^* \chi_j] = [\eta_i^* \xi_a] = \cdots = 0$. Here, $[\ldots]$ denotes average over the random variables. With these, $C(t)$ is given by:

$$C(t) = [(2\langle \xi\zeta|V|\eta\chi\rangle - \langle \xi\zeta|V|\chi\eta\rangle) \times \langle \eta(t)\chi(t)|V|\xi(t)\zeta(t)\rangle]. \qquad (5)$$

We consider in this work cases where the operation of $\hat{H}_0$ scales linearly with system size. This is true in semiempirical and basis-set calculations, where $\hat{H}_0$ is the sparse Fockian matrix (for sufficiently large systems) and it is true in DFT using grid representations (based on local/semi-local and hybrid exchange correlation potentials) where MP2 type calculations can be used in the context of double hybrid corrections.[23]

Naïve construction of random functions through Eqs. (4) requires calculation and storage of all possible eigenstates, a prohibitive task for large grid systems. Alternatively, one can perform this task using operator techniques. We first choose a function with random values at each grid point $x_n$, e.g. (for 3D systems),

$$\eta_0(x_n) = \frac{e^{i\theta_n}}{h^{3/2}}, \qquad (6)$$

where $\theta_n$ is a random sampled uniformly from $[0, 2\pi]$ and $h$ is the grid spacing. A similar procedure applies for constructing the other 3 random functions $\chi_0$, $\zeta_0$ and $\xi_0$. The random occupied-space function $\eta$ is now obtained by applying a "purification operator", $\eta(x) = \langle x|\theta(\mu - \hat{H}_0)|\eta_0\rangle$, where $\theta(E) \equiv \frac{1}{2}\text{erfc}\left(-\frac{E}{T}\right)$ and $\mu$ is the chemical potential (placed in the middle of the HOMO-LUMO gap). $T \approx \frac{1}{4}E_g$ is an artificial temperature and $E_g$ is the HOMO-LUMO gap. To apply $\theta(E)$ we use an iterative Chebyshev expansion for the step function:[33]

$$\eta(x) = \langle x|\theta(\mu - \hat{H}_0)|\eta_0\rangle = \sum_{k=0}^{K} \gamma_k \eta_k(x), \qquad (7)$$

where

$$\eta_1 = \hat{H}^N \eta_0, \qquad \eta_k = 2\hat{H}^N \eta_{k-1} - \eta_{k-2}, \qquad (8)$$

and $\hat{H}^N = \frac{\hat{H}_0 - \bar{H}}{\Delta H}$ is a normalized Hamiltonian[34] where $\bar{H}$ and $\Delta H$ are the average and half-width of the spectrum of $\hat{H}_0$; the normalization is required as the Chebyshev expansion applies for operators with spectra between -1 and 1.[34] The coefficients $\gamma_k$ are obtained by a numerical Fourier transform of the periodic function $f(\alpha) = \theta(\mu - (\bar{H} + \Delta H \cos \alpha))$. We repeat Eq. (7) also for $\chi$, $\zeta$ and $\xi$; however, for the latter two, since they must be projected into the virtual space, we subtract the "occupied" projected part from the original function to obtain the "virtual" projected part namely:

$$\zeta \leftarrow \zeta_0 - \zeta, \qquad \xi \leftarrow \xi_0 - \xi, \qquad (9)$$

The number of required Chebyshev steps in the purification stage can be estimated to be $K \approx 30 + \frac{10\Delta H}{T}$.

Consider now the time propagation of each random orbital. The total propagation time is $t_{max} \approx 3\tau$, determined by the parameter $\tau \gg E_g^{-1}$ governing the exponential damping of $C(t)$. Each random orbital is evolved using a Chebyshev series of the propagator[35] $e^{-i\hat{H}_0 t}$, and the number of terms in the series is $\sim 30 + t_{max}\Delta H$. Alternatively, one can use a split operator to carry out the propagation.

Summarizing the algorithm:

1. Construct the random complex functions, $\eta_0(x_n)$, $\chi_0(x_n)$, $\zeta_0(x_n)$, $\xi_0(x_n)$ on the grid (Eq.(4)).
2. Purify these functions to yield random occupied $\eta(x)$, $\chi(x)$ (Eqs.(7)-(8)) and virtual $\zeta(x)$, $\xi(x)$ (Eq. (9)) state functions.
3. Propagate the purify function to time $t_{max}$. At each time step along the propagation, evaluate the two-electron integrals and the MP2 correlation function given by Eq. (5), and carryout the time integration in Eq.(2).
4. Repeat steps 1-3 to average over the random functions.

To study the performance of this method, we first consider a Pariser–Parr–Pople Hamiltonian of a 1-dimensional system of $N$ sites and $N$ electrons at half-filling, described in ref. [36]. In the limit $N \to \infty$ this model has a gap $E_g = 0.11 E_h$ and energy span $\Delta H = 0.3 E_h$. For the calculation, we took (in atomic units) $\tau = 20$ and $t_{max} = 60$.

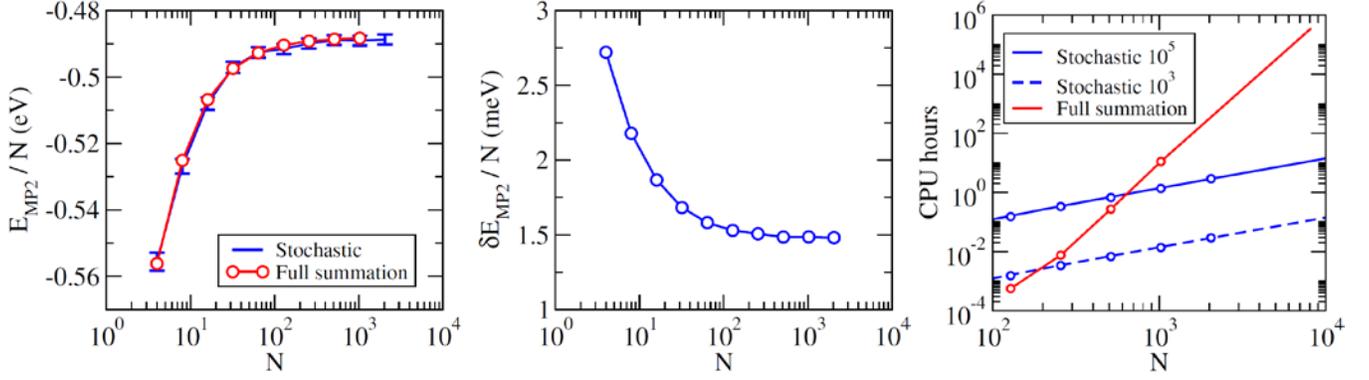

Figure 1 Left panel: Full summation vs. a stochastic calculation with ~$10^5$ iterations for the $E_{MP2}$ energy per electron for the half-filled model as a function of number of electrons (sites) $N$. Middle panel: The MC error bars per electron decrease towards a plateau as the system size grows. Right panel: The CPU times for full summation (scaling as $N^5$) vs. MC averaging (~$O(N)$) for the MP2 energy for $10^5$ (blue solid) and $10^3$ (blue dashed) MC runs. The costs above $N=1024$ for full-summation and $N=2048$ for the MC are extrapolated.

Figure 1 (left) shows the MP2 energies per site as a function of the number of sites (electrons) from a tiny ($N=4$) to a large system ($N=2048$). There is excellent agreement between the full-summation and stochastic results, and the scale of the deviation, per particle, is far better than chemical accuracy. The middle panel of Figure 1 shows that the statistical error per electron decreases towards a plateau as the number of electrons increases. The asymptotic value of the error is proportional to the square root number of MC iterations, and can be easily controlled. The right panel of Figure 1 shows the cost in CPU hours (measured per a regular 2.5 GHZ i7 Intel processor) of the full-summation and the stochastic methods (costs beyond 2048 sites for the stochastic method and 1024 for the full-summation are extrapolated based on the methods' scaling). We also show results for lower accuracy ($10^3$) MC runs. The calculations are very fast and still yield chemical accuracies of about $10-30$meV per electron. The stochastic method is already faster than the direct MP2 calculation for 600 sites at high accuracy and is two orders of magnitude faster for 2000 electrons/sites.

Table 1: Parameters for the hydrogen passivated silicon nanocrystals (NC). Shown, the number of silicon ($N_{Si}$) and hydrogen atoms ($N_H$), the number of electrons ($N_e$) the NC diameter ($D$), the number of grid-points and the occupied-virtual energy gap $E_g$.

| $N_{Si}$ | $N_H$ | $N_e$ | D (nm) | Grid | $E_g$(eV) |
|---|---|---|---|---|---|
| 1 | 4 | 12 |  | $8^3$ | 10.7 |
| 35 | 36 | 176 | 1.3 | $32^3$ | 3.9 |
| 87 | 76 | 424 | 1.6 | $48^3$ | 3.2 |
| 353 | 196 | 1488 | 2.4 | $64^3$ | 2.2 |
| 705 | 300 | 3120 | 3.0 | $72^3$ | 2.0 |

Next, we apply the stochastic method to hydrogen passivated spherical silicon nanocrystals (NCs) of several sizes, reaching systems with over 3000 electrons! The different systems are described in Table 1Table 1. We use a semi-empirical pseudopotential model to construct $\hat{H}_0$ as described in Ref. [37].

In the lower panel of Figure 2, we show the MP2 energy for the silicon NCs, which growth with the size of the NC. The full summation MP2 results can only be carried out for the smallest system, where we find that 20,000 MC iterations are sufficient to converge the results within an error smaller than 1meV.

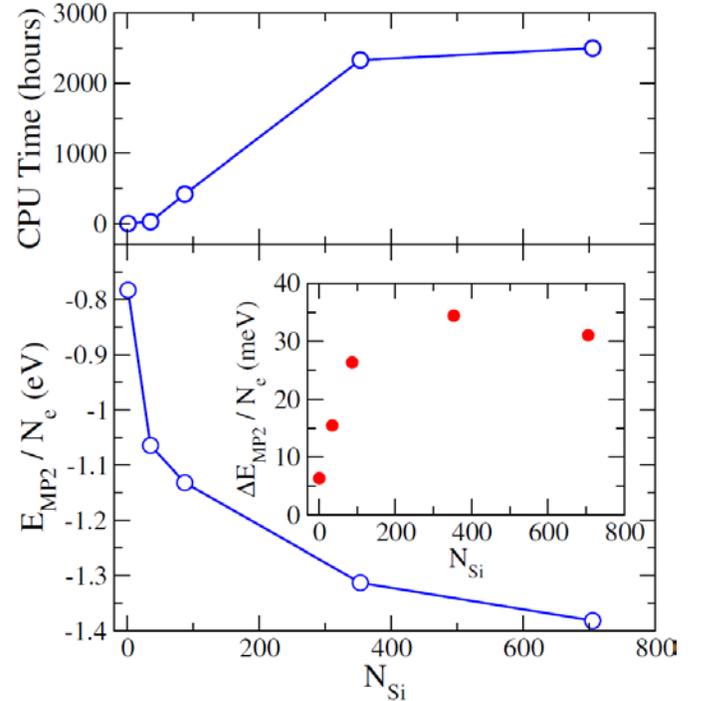

Figure 2: MP2 calculation data for hydrogen-passivated silicon nanocrystals (NCs) with $N_{Si}$ silicon atoms. Top panel: CPU time for computing the MP2 correlation energy to statistical error of 30meV. Bottom panel: The MP2 energy per electron as a function of NC size. Inset: the statistical error for 2500 MC iterations as a function of NC size.

The top panel of Figure 2 shows the total CPU time for a calculation that yields a statistical error of 30meV per electron. Of course, the error can be reduced by running longer in time and for the largest NC an error of 1% requires $10^4$ CPU hours. It is seen that CPU times grow by a factor of ~5 when going from the 87 to the 705 NCs. There are two distinct regimes for the scaling which are influenced by several factors: (a) The decrease of $E_g$ with NC size (almost a factor of 2 between the smallest and largest NCs studied) leads to an increase in the Chebyshev expansion by the same factor. (b) The statistical error grows when going from the small sized to 353 NCs, requiring more MC iterations. The difference in CPU time from $Si_{353}H_{196}$ to $Si_{705}H_{300}$ is rather small (sublinear), since the decrease in $E_g$ is rather small and this is also correlated with a decrease in the statistical errors. We note in passing that for even larger NCs (above 3nm), $E_g$ becomes nearly system size independent, and both the Chebyshev length and statistical errors saturate (the latter can also reduce with system size). Indication of this is seen in the inset of Figure 2, where the dependence of the statistical errors on system size (for 2500 iterations) is shown.

Summarizing, we have developed a stochastic method for calculating the MP2 correlation energies for systems of unprecedented sizes. The method was shown capable of addressing silicon nanocrystals with over 3000 electrons. Comparing to linear scaling local MP2[11,24] the present method is considerably less dependent on dimensionality since it is not highly sensitive to density matrix localization, as long as the zero order Hamiltonian can be operated in linear scaling. A second advantage of the method is its low memory requirements, involving only 4 orbitals, while linear scaling local MP2 will still require a huge memory in calculations for 3-dimensional dense systems. Another advantage is that the stochastic method is naturally parallelizable where each processor is an independent sampler and propagator and due to the low memory requirements can be implemented on graphic processing units.

There are several ways to extend the proposed approach. Higher order MP's, e.g., MP3 or MP4, seem within reach using similar ideas: one needs to replace energy denominators by time integrations as done here. This has the potential to allow efficient calculation of high order perturbation energy calculations. Another interesting feature is the extension to finite temperatures, which is straightforward to implement within the present formalism. Finally, we plan to investigate whether for very large sizes a localization scheme could reduce the error per electron as the system gets larger.

DN is grateful to Julien Toulouse who suggested applying our exchange scheme also to MP2 energies and to Andreas Savin for a useful discussion. DN was supported by the NSF, grant CHE-1112500. RB and DN were supported by the US-Israel Binational Foundation (BSF). ER would like to thank the Israel Science Foundation (grant number 611/11) for financial support.